\documentclass[conference, letterpaper, twocolumn]{IEEEtran}

\IEEEoverridecommandlockouts
\usepackage{graphicx, epsfig, amssymb, amsmath, amsthm}
\usepackage{amsmath, amssymb, amsfonts}
\usepackage{slashbox, pict2e}
\usepackage{multirow}
\usepackage{xcolor}
\usepackage{times}
\usepackage{cases}
\usepackage{array}
\usepackage{float}
\usepackage{algorithm}
\usepackage{algorithmic}
\usepackage{cite}
\usepackage{bm}
\usepackage{setspace}
\usepackage{textcomp}
\usepackage{longtable}
\usepackage{makecell}
\usepackage{subfigure}
\usepackage{siunitx} 
\usepackage{flushend} 
\usepackage{epsfig, epstopdf}

\begin{document}
\title{System-Level Simulation Framework for NB-IoT: Key Features and Performance Evaluation}

\bibliographystyle{IEEEtran}
\newtheorem{lemma}{Lemma}

\author{Shutao~Zhang, Wenkun Wen, Peiran~Wu, Hongqing~Huang, Liya~Zhu, Yijia~Guo, Tingting Yang, and Minghua~Xia 
\thanks{S. Zhang, P. Wu, H. Huang, L. Zhu, Y. Guo, and M. Xia are with the School of Electronics and Information Technology, Sun Yat-sen University, Guangzhou, 510006, China (e-mail:\{zhangsht3, huanghq29, zhuly55, guoyj33\}@mail2.sysu.edu.cn, \{wupr3, xiamingh\}@mail.sysu.edu.cn).}
\thanks{W. Wen is with the Guangzhou Techphant Co. Ltd. Guangzhou, 510310, China (e-mail: wenwenkun@techphant.net)}
\thanks{T. Yang is with Peng Cheng Laboratory, Shenzhen, China, and the Navigation College, Dalian Maritime University, Dalian 116026, China (Email: yangtingting820523@163.com).}
}

\maketitle

\begin{abstract}
\noindent  Narrowband Internet of Things (NB-IoT) is a technology specifically designated by the 3rd Generation Partnership Project (3GPP) to meet the explosive demand for massive machine-type communications (mMTC), and it is evolving to RedCap. Industrial companies have increasingly adopted NB-IoT as the solution for mMTC due to its lightweight design and comprehensive technical specifications released by 3GPP. This paper presents a system-level simulation framework for NB-IoT networks to evaluate their performance. The system-level simulator is structured into four parts: initialization, pre-generation, main simulation loop, and post-processing. Additionally, three essential features are investigated to enhance coverage, support massive connections, and ensure low power consumption, respectively. Simulation results demonstrate that the cumulative distribution function curves of the signal-to-interference-and-noise ratio fully comply with industrial standards. Furthermore, the throughput performance explains how NB-IoT networks realize massive connections at the cost of data rate. This work highlights its practical utility and paves the way for developing NB-IoT networks.
\end{abstract}


\section{Introduction}
\label{Section-Introduction}

With the rapid development of the Internet of Things (IoT), many companies and organizations have launched their designs of IoT communication systems. For example, extended coverage-GSM (EC-GSM), enhanced machine-type communications (eMTC), and narrowband Internet of Things (NB-IoT) are based on the Long Term Evolution (LTE) cellular communication system that works in licensed frequency bands \cite{8377168, 8904267, 7419993, 7876968}. At the same time, LoRa and Sigfox are two typical IoT techniques that work in unlicensed frequency bands \cite{8067462, 8903531, 8742582}. Compared with EC-GSM and eMTC, NB-IoT has a new air interface design with a $3.75$ or $15$ kHz subcarrier in a $180$ kHz bandwidth, which is more suitable for the massive IoT scenario. Compared with LoRa and Sigfox, the precise standard of NB-IoT is non-proprietary, and the corresponding specifications about the physical layer and higher layer are publicly available from the 3rd Generation Partnership Project (3GPP). As a result, NB-IoT is the most promising low power and wide area (LPWA) technique among the technologies mentioned above \cite{NBImplementations, NBEvolutions, NBSimulation}. Although the fifth generation (5G) mobile communications was commercialized at the end of 2019, the massive machine-type communication (mMTC) standard was finally fulfilled on July 9, 2020. On July 9, 2020, the International Telecommunication Union Radiocommunication Sector (ITU-R) formally approved 5G technology as IMT-2020 5G standard at the ITU-R working party 5D (WP5D) \#35 meeting and thus, NB-IoT becomes essential to the mMTC applications. In 3GPP Release 17 forward, the NB-IoT is evolving to RedCap to allow a transition from massive to broadband IoT connectivity \cite{singh2024towards}.

\begin{figure}[t]
	\centering
	\begin{subfigure}[{\scriptsize The architecture of NB-IoT networks.}]{
			\includegraphics [width=3.5in, clip, keepaspectratio]{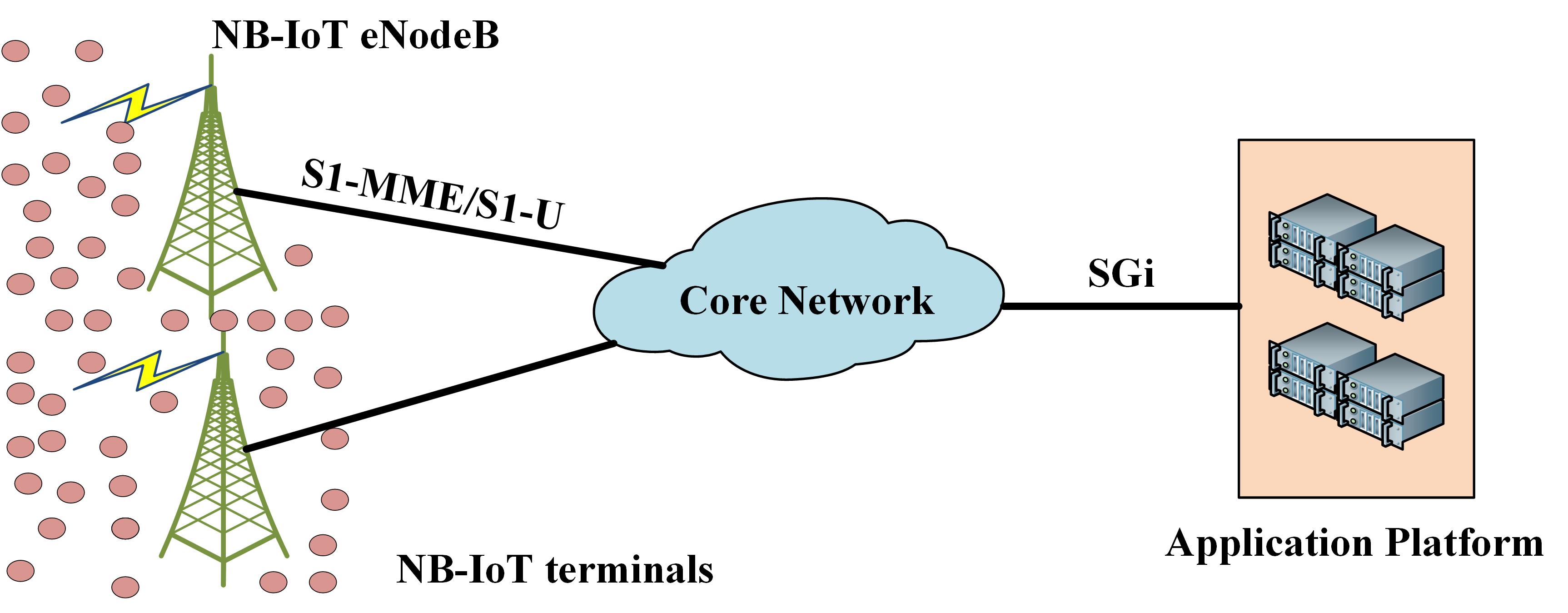}
			\label{Architecture}}
	\end{subfigure}
	\\
	\begin{subfigure}[{\scriptsize  The entities of  NB-IoT networks.}]{
			\includegraphics [width=3.5in, clip, keepaspectratio]{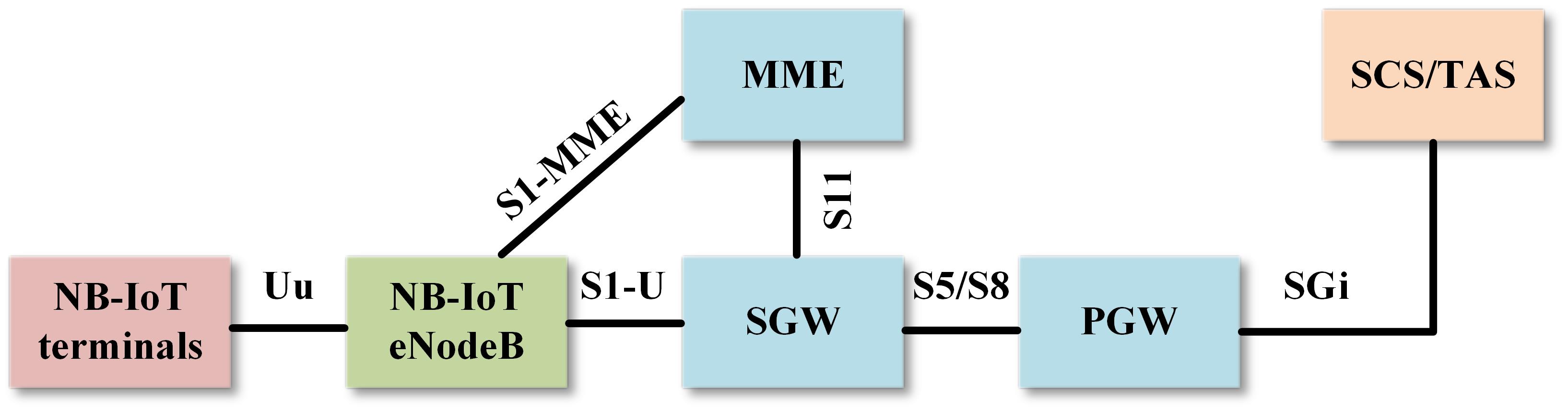}
			\label{Entity}}		
	\end{subfigure}
    \vspace{-2pt}
    \caption{The NB-IoT network architecture and functional entities.}
    \vspace{-2pt}
\end{figure}

Figures~\ref{Architecture} and \ref{Entity} depict the NB-IoT network architecture and corresponding functional entities, respectively. It is seen from Figure~\ref{Architecture} that the NB-IoT network includes the NB-IoT terminal, NB-IoT eNodeB, core network, and application platform. For low-cost and power saving, the hardware resource of the NB-IoT terminal is strictly limited. NB-IoT eNodeB serves as a base station to access massive NB-IoT terminals. NB-IoT terminal and NB-IoT eNodeB can be deployed in stand-alone, guard-band, and in-band modes. 

On the other hand, Figure~\ref{Entity} shows that the NB-IoT core network, derived from the evolved packet core (EPC) in the LTE system, consists of three functional entities: mobility management entity (MME), serving gateway (SGW), and public data network gateway (PGW). Among them, MME works for session-related management and signaling control; SGW routes and forwards local data packets, and PGW is responsible for external data connection, such as packet filtering and IP address allocation. The NB-IoT application platform comprises a service capability server (SCS) and a third-party application server (TAS). As for the interface of network entities, the Uu interface refers to the communication rules between the NB-IoT terminal and NB-IoT eNodeB. The S1-MME interface bears signaling interaction between  NB-IoT eNodeB and  MME, while the S1-U interface bears data transmission between NB-IoT eNodeB and  SGW. The S11 interface connects MME and SGW, with the S5 and/or S8 interface exchanging information between SGW and PGW. The SGi interface enables communication from PGW to outside servers in application platforms like SCS/TAS to support various NB-IoT services. 

When deploying an NB-IoT network in practice, the link-level and system-level performance evaluation is instrumental and crucial. The link-level performance evaluation focuses on the physical layer of NB-IoT, i.e., an end-to-end link \cite{9194757, lagen2020new, iiyambo2024survey}. In the uplink of NB-IoT, the narrowband physical random access channel (NPRACH) and a narrowband physical uplink shared channel (NPUSCH) were both modeled in \cite{8922625}. The design details and performance of NPRACH and NPUSCH were also shown in \cite{9079923} and \cite{Aoxiang}, respectively. In the downlink of NB-IoT, three physical channels were newly introduced, namely, narrowband physical broadcast channel (NPBCH), narrowband physical downlink control channel (NPDCCH), and narrowband physical downlink sharing channel (NPDSCH). An efficient downlink receiver for NB-IoT was designed in \cite{Zhan2004}. Moreover, the synchronization signals, consisting of narrowband primary synchronization signal (NPSS) and narrowband secondary synchronization signal (NSSS), were mainly used for time and frequency synchronization between NB-IoT terminals and associated NB-IoT eNodeB \cite{npssnsss, Ali7918505, yu2023energy}. Our previous work \cite{9079923, Aoxiang, Zhan2004} has constructed the uplink and downlink receiver for NB-IoT conforming to Release 15, and link-level performance satisfies the benchmarks designated by 3GPP \cite{36101, 36104}.  

On the contrary, the system-level performance evaluation focuses on the overall system performance of an NB-IoT network through a multi-cell and massive-terminal network topology. In the literature, for instance, channel utilization was defined in \cite{NBSimulation} to simulate the NB-IoT network. However, this work presented the NB-IoT model based on the OPNET platform without a specific system-level simulation process. As for radio resource scheduling in the NB-IoT network, a dual-layer link adaptation scheme was proposed in \cite{7842562} by first introducing repetition number to NB-IoT. An iterative algorithm based on cooperative approaches was developed in the subcarrier and power allocation of the NB-IoT network to maximize achievable data rate  \cite{8306882}. The authors noticed that NB-IoT network performance evaluations were typically done separately at the link and system levels. This approach led to incomplete evaluations and potentially risky engineering practices. Consequently, our research focuses on designing an efficient system-level simulation platform to evaluate the increasingly complex NB-IoT network better.

In this paper, we design a comprehensive NB-IoT system-level simulation framework based on object-oriented programming in the MATLAB platform with version R2020a. The comprehensive system-level simulation framework contains both link- and system-level simulators. To make a tradeoff between low complexity and high accuracy, we adopt the approach of quasi-dynamic system-level simulation, where the quick network layout in static system-level simulation and high-accuracy modeling in dynamic system-level simulation are exploited simultaneously. To further analyze and verify the key features of the NB-IoT network, we evaluate the maximum coupling loss (MCL), traffic model, and extended discontinuous reception (eDRX) to evaluate the capabilities of the enhanced coverage, massive connections, and low-power consumption, respectively. Finally, we discuss the cumulative distribution function (CDF) of the signal-to-interference-and-noise ratio (SINR) and prove that the NB-IoT network supports massive connections of NB-IoT terminals at the cost of the normalized user throughput.

The rest of this paper is organized as follows. Section~\ref{Section-llss} presents the structure of the link-level simulator, and Section~\ref{Section-slss} describes the structure of the system-level one. Section~\ref{Section-Simulation} provides detailed insights into three key technologies of the NB-IoT network. Section~\ref{Section-Results} discusses and analyzes the simulation results, and finally, Section~\ref{Section-Conclusions} concludes the paper.

\begin{figure*}[!t]
	\begin{center}
		\includegraphics[width=7in, clip, keepaspectratio]{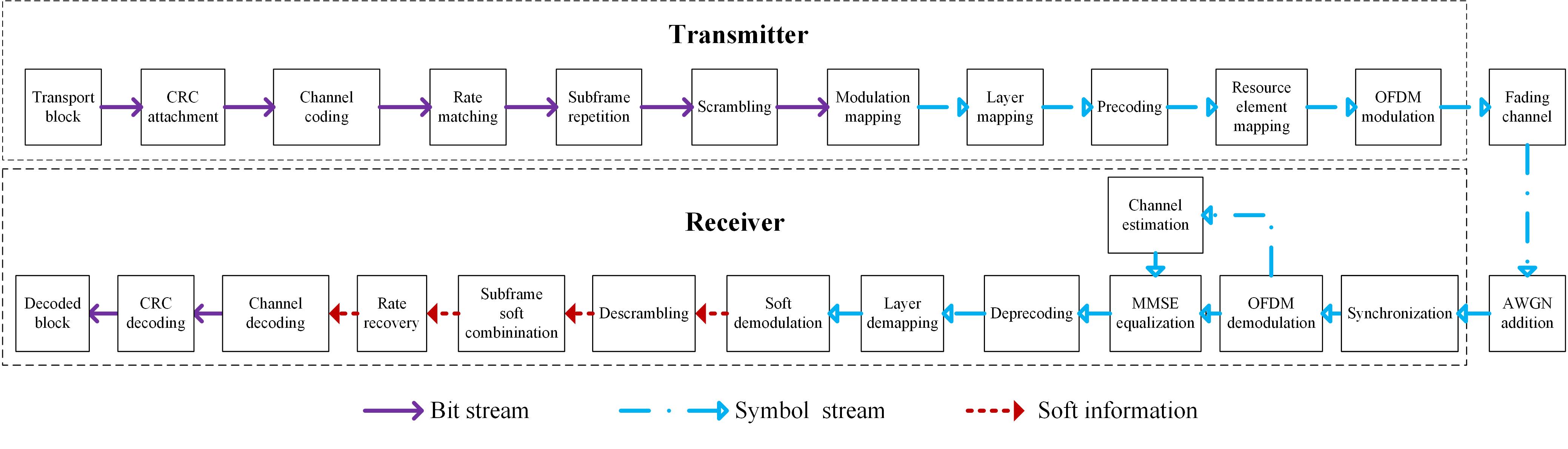}
		\vspace{-2pt}
		\caption{The transceiver chain of the link-level simulator.}
		\label{Fig-1}
	\end{center}
	\vspace{-2pt}
\end{figure*}

\section{Link-Level Simulator Structure} \label{Section-llss}

The link-level simulator of NB-IoT focuses on the physical layer issues, such as channel coding, modulation, and equalization. As illustrated in Figure~\ref{Fig-1}, the transceiver mainly describes the process of encoding specified in the 3GPP Technical Specification (TS) 36.212 \cite{36212}, and the process of modulation specified in the 3GPP TS 36.211 \cite{36211}. Given some data blocks to be transmitted, the cyclic redundancy check  (CRC) bits are calculated and appended to the end of each transport block for error check. Then, channel coding and channel interleaving are applied for error correction. The block of rate matching adjusts the number of coded bits to the scheduled physical resources of one subframe, followed by subframe repetition, a key feature of NB-IoT for expanded coverage and enhanced transmission reliability. Next, the downlink scrambling benefits identifying cells and breaking continuous bit sequences of $0$s or $1$s. Afterward, the scrambled bits are modulated by the quadrature phase shift keying (QPSK) constellation, followed by layer mapping and precoding, and then the modulated symbols are mapped to different physical resources through resource element (RE) mapping. Finally, the orthogonal frequency division multiplexing (OFDM) modulation is applied, after which the symbols are transmitted from the antenna(s).

At the receiver side, the received signals attenuated by the fading channel are corrupted by independent identically distributed (i.i.d.) additive white Gaussian noise (AWGN). In principle, the transmitted signals can be recovered by using an inverse process to that at the transmitter. To this end, synchronization is first acquired using NPSS, and then physical cell identification is obtained using NSSS. After OFDM demodulation, channel estimation is performed to obtain channel frequency response (CFR), and then channel equalization is performed with the help of a narrowband reference signal (NRS). Next, QPSK demodulation provides soft information that contains the likelihood of received bits and is extremely useful for subsequent channel decoding. Afterward, a soft combination is performed to exploit the subframe repetitions. Finally, rate recovery, channel decoding, and CRC decoding are performed sequentially.

\begin{figure*}[!t]
	\begin{center}
		\includegraphics[width=5.0 in, clip, keepaspectratio]{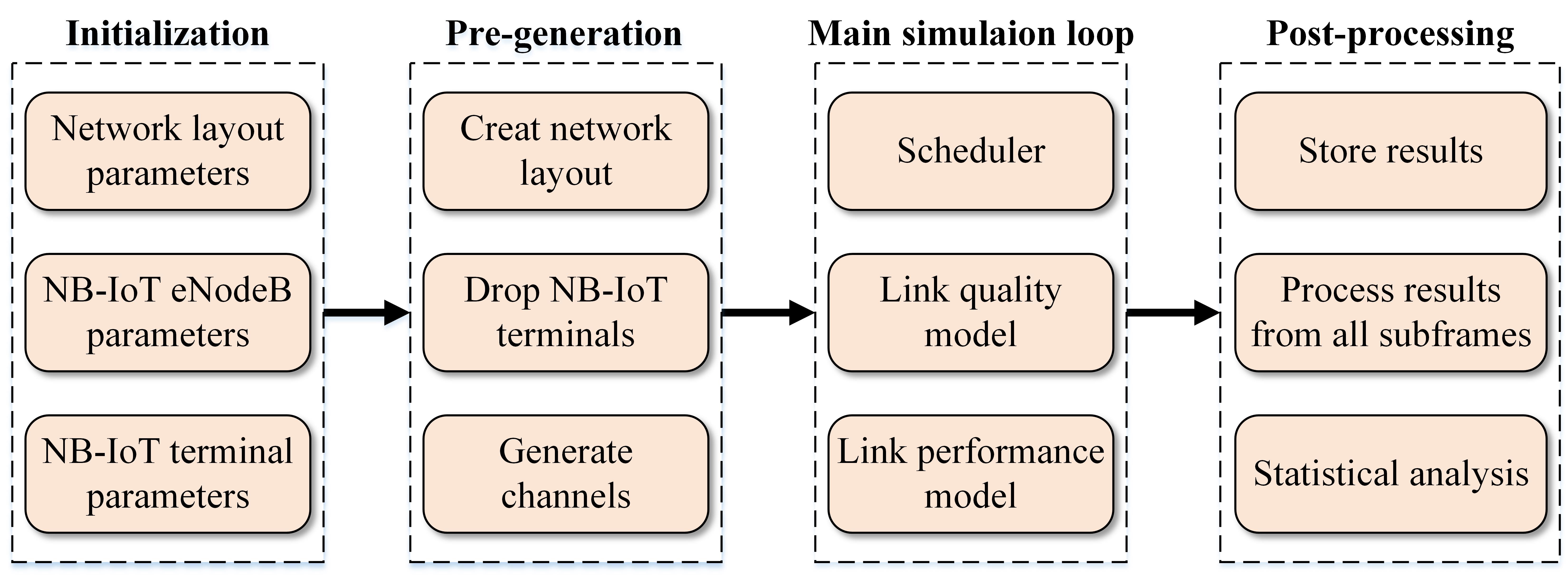}
		\caption{The building blocks of the system-level simulator.}
		\label{SystemModel}
	\end{center}
\end{figure*}

\section{Design of System-Level Simulator} \label{Section-slss}

As shown in Figure~\ref{SystemModel}, the system-level simulator creates an efficient signal processing flow consisting of four main blocks: initialization, pre-generation, main simulation loop, and post-processing. The initialization block configures network topology parameters, NB-IoT eNodeB parameters, and NB-IoT terminal parameters. The pre-generation block creates network topology, places NB-IoT terminals, and generates propagation channels. The main simulation loop includes the scheduler, link quality, and link performance analysis models. Based on the link-level simulation results, the system-level simulator abstracts the physical layer of NB-IoT into the link quality and performance models. It simulates the system performance of NB-IoT in a multi-cell and massive-terminal network topology. The last post-processing block stores the simulation results combines related simulation results in different time subframes and performs statistical analysis. Next, we discuss the details of each building block. 

\subsection{Initialization}
To start with any system-level simulation experiments, we must initially configure all the basic parameters for the network layout, NB-IoT eNodeB, and NB-IoT terminals. The parameters of the network layout consist of the number of cells, the number of sectors, and the distance between NB-IoT eNodeBs. The parameters of NB-IoT eNodeB and terminals include antenna gain, transmit power and frequency band. The parameter setting mentioned above follows the standard ones defined in the 3GPP TS 36.888 \cite{36888}. Specifically, we have structured the network layout as a classic three-layer cellular structure with $19$ NB-IoT eNodeBs, each covering $3$ sectors. The distance between two adjacent NB-IoT eNodeBs is precisely $1732$ meters. The antenna gain and transmit power of NB-IoT eNodeB are set at $18$ dBi and $43$ dBm, respectively. Similarly, the NB-IoT terminal's antenna gain and transmit power are $-4$ dBi and $23$ dBm, respectively. The RF center frequency is fixed at $900$ MHz. Based on these settings, the system-level simulation runs in every transmission time interval (TTI) of 1 millisecond per subframe. During every TTI, a codeword is formulated and transmitted. 
%{\textcolor{red} {(We must cite an authoritative source of these parameters, like 3GPP TS; otherwise, their values will be challenged.)}}

\begin{figure*}[t!]
	\centering
	\begin{subfigure}[{\scriptsize The three-layer network layout.}]{
			\includegraphics[height=2.5in]{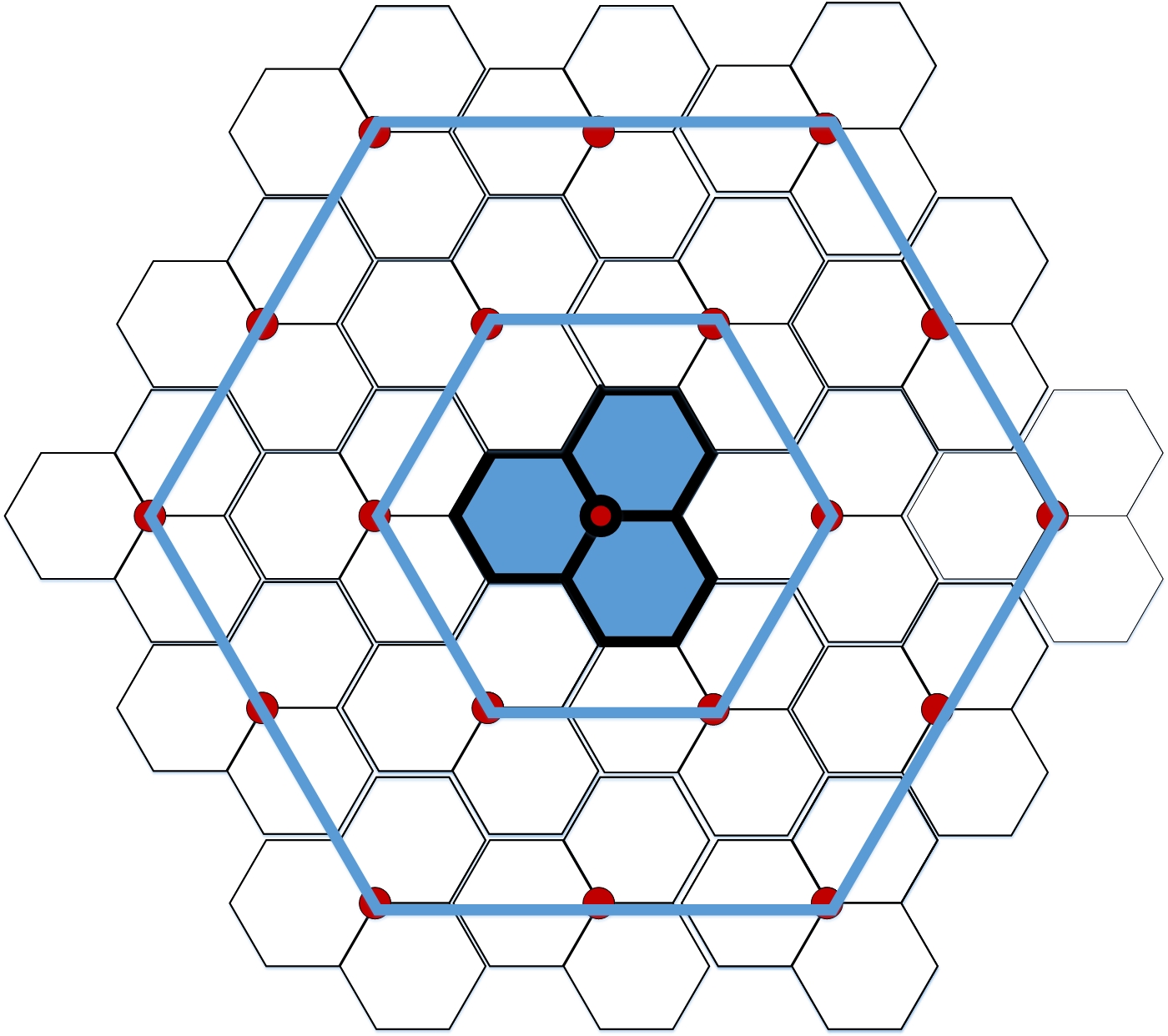}
			\label{network}}
	\end{subfigure}
	\qquad
	\begin{subfigure}[{\scriptsize NB-IoT eNodeBs (red solid spots) and terminals (blue circles).}]{
			\includegraphics[height=2.5in]{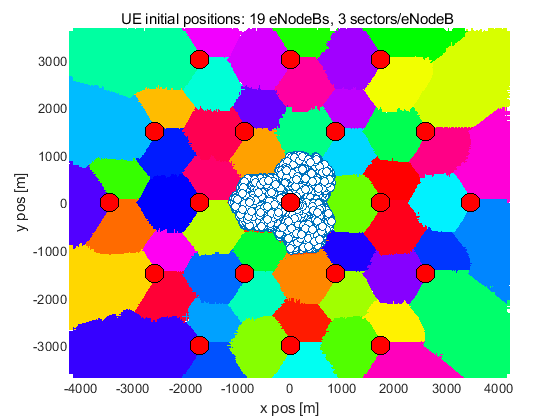}
			\label{eNodeB_point2}}
	\end{subfigure}
	\caption{The network model of the system-level simulator.}
	\label{Fig-4}
\end{figure*}

\subsection{Pre-generation}
The pre-generation block creates the network topology, places NB-IoT terminals, and generates the propagation channels. The network layout determines the distribution of NB-IoT eNodeBs and the location of massive NB-IoT terminals. Once the network layout is set, propagation channels between NB-IoT eNodeBs and NB-IoT terminals can be generated, taking into account large-scale path loss, shadowing effect, and small-scale fading parameters.

%The network layout of NB-IoT consists of $19$ NB-IoT cells, each comprising $3$ hexagonal sectors with an NB-IoT eNodeB located at its center. As shown in Figure~\ref{network},  there are $12$ NB-IoT eNodeBs in the outermost layer, $6$ NB-IoT eNodeBs in the middle layer, and $1$ NB-IoT eNodeB in the innermost layer. The system-level simulation focuses on the innermost NB-IoT eNodeB with three sectors highlighted in blue, and the other $18$ NB-IoT eNodeBs work as interfering ones. After setting the network layout, we place NB-IoT terminals in the region of interest (ROI). By setting the pixel resolution to 5 m, a two-dimensional pixel map can be created according to the network layout. Then, each pixel can be placed with an NB-IoT terminal. As shown in Figure~\ref{eNodeB_point2}, NB-IoT eNodeBs and terminals are highlighted with red solid points and blue dots, respectively. Thousands of NB-IoT terminals with uniform distribution are dropped in the ROI randomly, which generally stay still or have slow-speed movement.

As illustrated in Figure~\ref{network}, the NB-IoT network layout consists of $19$ NB-IoT cells, each with $3$ hexagonal sectors and an NB-IoT eNodeB at its center. There are $12$ NB-IoT eNodeBs in the outermost layer, $6$ in the middle layer, and $1$ in the innermost layer, denoted by red solid spots in the figure. The system-level simulation focuses on the innermost NB-IoT eNodeB, with $3$ sectors highlighted in blue, while the other $18$ NB-IoT eNodeBs act as interfering ones. Once the network layout is set, NB-IoT terminals are placed in the region of interest (ROI) using a two-dimensional pixel map with a pixel resolution of $5$ meters. Each pixel represents the location of an NB-IoT terminal. Thousands of NB-IoT terminals, evenly distributed in the ROI, are then randomly positioned, shown in blue circles in Figure~\ref{eNodeB_point2}, and they generally have limited movement or stay stationary.

The propagation channels need to be pre-generated and stored before entering the main simulation loop. The channel generates large-scale fading (including path loss and shadowing effect) and small-scale fading, which can be expressed as
\begin{equation} \label{Eq_TRx}
P_{\rm Rx}(t, n) = P_{\rm Tx}(t, n) + \alpha + \beta + F(t, n), 
\end{equation}
where $P_{\rm Rx}(t, n) $ and $P_{\rm Tx}(t, n)$ denote the received power at the eNodeB and transmit power of a terminal at time instant $t$ on the subcarrier $n$, in the unit of \si{dBm}, respectively; $\alpha$, $\beta$ and $F(t, n)$ represent the path loss, shadowing effect and small-scale fading, in the unit of \si{dB}, respectively. By accounting for the multipath effect, the small-scale fading $F(t, n)$ at time instant $t$ on the subcarrier $n$ can be determined by its power delay profile \cite{36101}. 

The path loss $\alpha$ in \eqref{Eq_TRx} can be explicitly computed as 
\begin{equation}\label{Eq4-1}
\alpha = \max\left( L - G_{ \rm BS} - G_{\rm terminal}, \ \Psi \right) + \delta + \xi,
\end{equation}
where the value of $L$ depends on the path loss model of a city or suburb, determined as \cite{36942}
\begin{align}\label{Eq_Path_Loss}
L =  & \ 40 \times \left(1 - 4 \times 10^{-3}  H\right)\log_{10}R - 18\log_{10}H \nonumber \\ & + 21  \log_{10}f + 80 \ \text{dB},
\end{align}
in which $H = 15$ \si{m} stants for the antenna height of NB-IoT eNodeB, $R$ is the distance between NB-IoT eNodeB and an associated terminal, and $f = 900$ \si{MHz} denotes the carrier frequency. With these parameter fixed, \eqref{Eq_Path_Loss} reduces to
\begin{equation}\label{Eq-MCL}
L = 120.9 + 37.6  \log_{10}R \ \text{dB}.
\end{equation}
On the other hand, the paramerter $G_{ \rm BS}$ in \eqref{Eq4-1} denotes the antenna gain of NB-IoT eNodeB \cite{45914}, given by
\begin{equation}
G_{ \rm BS} = -\min\left(12\left(\frac{\theta}{65^{\circ}}\right)^2, \ 23 \right) + G_{ \rm max},
\end{equation}
where $ \theta \in \left[ -180^{\circ},  180^{\circ} \right]$ counts the angle between NB-IoT eNodeB and an associated terminal in the vertical plane; $G_{ \rm max} = 18$ dBi is the maximum antenna gain of NB-IoT eNodeB. Finally, $G_{\rm terminal}$ in \eqref{Eq4-1} is set to $-4$ \si{dBi} due to the isotropic antenna of the NB-IoT terminal; $\Psi \approx 178.96$ \si{dB} denotes the minimum path loss according to \eqref{Eq-MCL} assuming that the minimum distance between NB-IoT eNodeB and a terminal is $35$ \si{m}; $\delta = 3$ \si{dB} stands for the cable loss of NB-IoT eNodeB, and $ \xi = 20$ \si{dB} is the building penetration loss.

The value of the shadowing parameter $\beta$ in \eqref{Eq_TRx} can be determined as follows. In principle, the shadowing effect introduced by large obstacles in the propagation path leads to large-scale fading fluctuations. In mathematics, it is modeled as a log-normal distribution of mean $0$ dB.  To generate spatially correlated shadowing, we adopt the method designed in \cite{1651489}. Specifically, let $x$ denote the distance from an NB-IoT eNodeB to a point in space, then the correlation coefficient is given by
\begin{equation} \label{Eq-correlation_function}
	r(x) = e^{-\tau}x, \ x> 0,
\end{equation}
where  $\tau$ is the normalized factor to be determined by the environment (e.g., a value of $\tau  = 1/20 $ is suggested for urban scenarios). Given an $m \times n$ pixel map, the correlation matrix $\bm{R}\in \mathbb{R}^{mn \times mn}$ can be calculated by using \eqref{Eq-correlation_function}. Then, performing Cholesky decomposition on $\bm{R}$ gives $\bm R = \bm{LL}^T$. As a result, the shadowing random vector $\bm s \in \mathbb{R}^{mn}$ is computed as
\begin{equation}
\label{Eq-correlation_value}
\bm s = \bm{L}\bm{a}, 
\end{equation}
where $\bm{a} = \left[a_1, a_2, \cdots, a_{mn}\right]^{T}$ is the random vector satisfing $\mathbb{E}\left[ \bm{aa}^{H} \right] = \bm{I}$. Finally, the parameter $\beta$ in \eqref{Eq_TRx} can take any element in $\bm{s}$, representing the corresponding pixel's shadowing value in the network layout.

\subsection{Main Simulation Loop}
\begin{figure}[t]
	\begin{center}
		\includegraphics[width=1.25in]{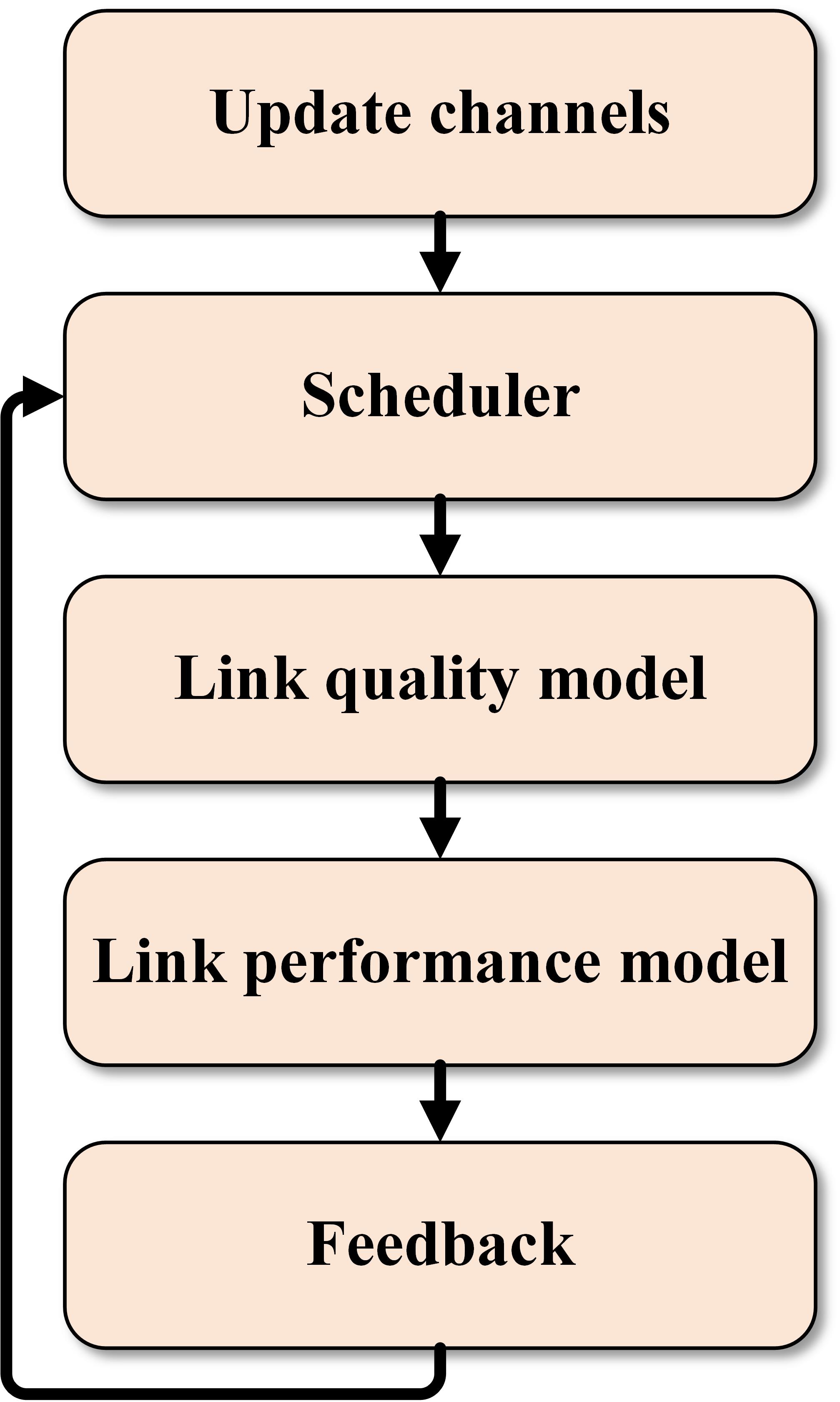}
		\caption{The main simulation loop of the system-level simulator.}
		\label{mainloop}
	\vspace{-10pt}
	\end{center}
\end{figure}

Figure~\ref{mainloop} shows the five steps performed in the main simulation loop for each TTI iteration: updating the channel parameters regarding the small-scale fading, scheduling available time-frequency resources to NB-IoT terminals, calculating the effective SINR in the link quality model, judging whether transport block is decoded correctly and mapping SINR to channel quality indicator (CQI) in the link performance model, and finally, sending feedback to the scheduler. Next, we elaborate on the technical issues in each step.

\subsubsection{Update Channels}
A system-level simulation experiment generally lasts thousands of TTIs, equivalent to several seconds in a real-world network. Although the large-scale fading is time-varying, it varies slowly and can be assumed to remain unchanged during these few seconds. Therefore, the configurations of large-scale fading parameters, such as path loss and shadowing, need not be updated. Meanwhile, the small-scale fading channels are updated in consecutive TTIs according to the pre-generated channel model.

\subsubsection{Scheduler}
After a delay in transmission, the scheduler receives feedback on CQI from NB-IoT eNodeB in the uplink or NB-IoT terminals in the downlink. In NB-IoT, the CQI value ranges from $0$ to $13$, with higher values indicating better channel quality. Upon receiving CQI, the scheduler must determine the modulation and coding scheme (MCS), the number of repetitions, and the transmission format. In NB-IoT, the MCS value equals the value of CQI, and the transmission block size can be obtained through MCS. 

The uplink transmission format is based on the resource unit (RU), composed of a $3.75/15$ kHz subcarrier and corresponding time slot. In contrast, the downlink transmission format is based on $12$ subcarriers and $14$ OFDM symbols. The scheduler uses the classic round-robin (RR) and proportional fair (PF) algorithms for scheduling. The RR algorithm allows NB-IoT terminals to share time-frequency resources regardless of instantaneous channel quality. In contrast, the PF algorithm considers fairness and efficiency based on the priority of the NB-IoT terminal. The terminal with the best channel quality has the highest priority and is scheduled first. If a terminal has poor channel quality and has not been scheduled for a long time, its priority will gradually increase until it is scheduled.

The scheduler must also handle the hybrid automatic repeat request (HARQ) process to allow for repetitive transmissions. In the uplink HARQ process, the new data indication (NDI) is used to signal whether an NB-IoT terminal can continue to send new data. If the transport block decodes the received codeword correctly, the NB-IoT eNodeB generates NDI feedback to inform the scheduler that the NB-IoT terminal should transmit new data. If the transport block is not decoded correctly, the NB-IoT eNodeB does not generate NDI. When the NB-IoT terminal does not receive NDI within a certain period, it retransmits the previous transport block. For the downlink HARQ process, the ACK/NACK mechanism is applied to indicate whether the decoding result is correct or incorrect. The NB-IoT terminal sends the ACK/NACK information to the scheduler as feedback.

\subsubsection{Link Quality Model}
Simulating the complete physical layer for all links between an NB-IoT eNodeB and its associated massive terminals is often challenging due to the high time complexity and extensive storage memory requirements. To expedite the simulation, we abstract the physical layer as a link quality model and a performance analysis model \cite{4gc, 5g} in the developed simulator. In the link quality model, the received SINR of a subcarrier can be calculated using a pre-generated channel model. Specifically, in the single-input single-output (SISO) transmission mode of NB-IoT, the received SINR of the $i^{\rm th}$ subcarrier can be calculated as:
\begin{equation} \label{SINR_subcarrier}
	\gamma_i = \frac{\left|h_{0}^{ (i)}\right|^{2}P_{{\rm Tx}, 0}^{\ (i)}}{\sum\limits_{k=1}^{N} \left|h_{k}^{(i)}\right|^{2} P_{{\rm Tx}, k}^{\ (i)} + \sigma^{2}_i},
\end{equation}
where $P_{{\rm Tx}, k}^{\ (i)}$ and $|h_{k}^{ (i)}|^{2}$ denote the transmit power and the channel gain from the $i^{\rm th}$ subcarrier of the $k^{\rm th}$ cell to its associated eNodeB, respectively. The $0^{\rm th}$ cell is the target cell located at the center of the network. The index $k = 1, \cdots, N$ denotes the interference cells surrounding the target cell, and in this case $N = 18$. Also, $\sigma_{i}^{2} $ in \eqref{SINR_subcarrier} represents the receiver noise power corresponding to the $i^{\rm th}$ subcarrier of the $0^{\rm th}$ cell. 

Based on the per-subcarrier SINR  computed by \eqref{SINR_subcarrier}, the effective SINR for an NB-IoT terminal can be calculated by the method of exponential effective signal-to-interference-and-noise ratio mapping (EESM) \cite{4735883}. Here, a set of per-subcarrier SINR $\bm{ \gamma} = [\gamma_1,\cdots,\gamma_{P}]^{\rm T}$ assigned to the NB-IoT terminal is mapped into an effective SINR $ \gamma_{\rm eff} $, defined as
\begin{equation} \label{SINReff}
	\gamma_{\rm eff} = {\rm EESM}(\bm{\gamma}, \eta) = -\eta\ln\left[ \frac{1}{ P}\sum_{i = 1}^{ P}\exp\left( \frac{\gamma_i}{\eta} \right)\right],
\end{equation}
where $P$ denotes the number of subcarriers assigned to the NB-IoT terminal, and $\eta$ is the factor regarding the modulation scheme. For intance, $\eta = 2$ for the QPSK modulation \cite{4735883}.

\begin{figure*}[t!]
	\centering
	\begin{subfigure}[{\scriptsize BLER curves of link performance model.}]{
			\includegraphics[height=3.0in]{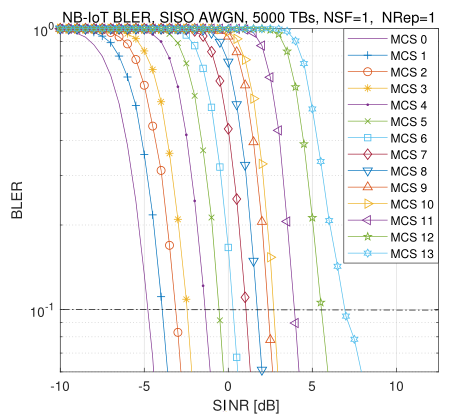}
			\label{BLER}}
	\end{subfigure}
	\qquad
	\begin{subfigure}[{\scriptsize CQI mapping with BLER = $10\%$.}]{
			\includegraphics[height=3.0in]{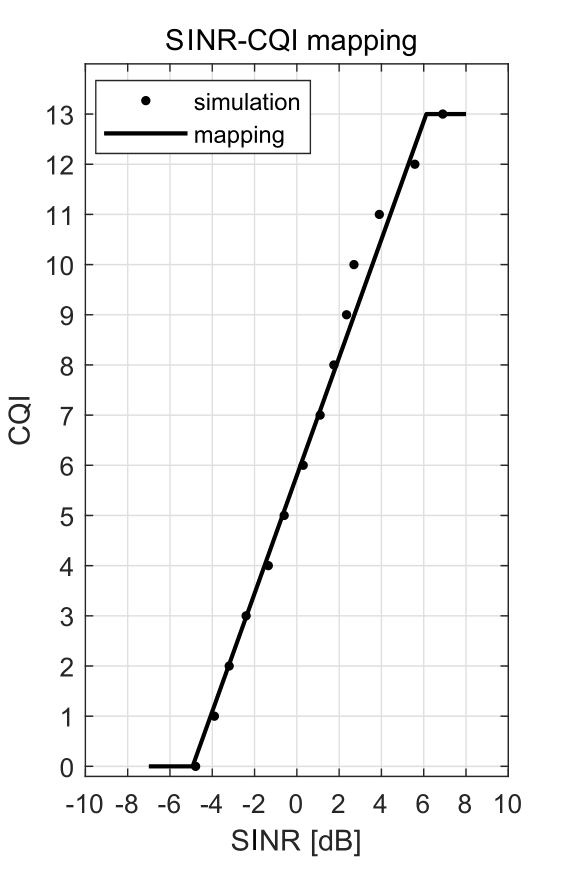}
			\label{mapping}}
	\end{subfigure}
	\caption{Physical-layer mapping used in the system-level simulator.}
\end{figure*}

\subsubsection{Link Performance Model}
The effective SINR ($\gamma_{\rm eff}$) calculated in the link quality model is used as input into the link performance model to obtain the decoding result of the transport block and CQI. The NB-IoT network has $14$ kinds of MCSs and $14$ relevant CQIs, as depicted in Figures~\ref{BLER} and \ref{mapping}. Once the number of subframes ($N_{\rm SF}$) and MCS are known, the transport block size (TBS) can be determined as specified in Subclause 8.6 of \cite{36213}. For example, Figures~\ref{BLER} and \ref{mapping} show the simulation results for SISO configuration with a bandwidth of $180$ kHz in an AWGN channel, with $N_{\rm SF} = 1$ and the number of repetitions $N_{\rm Rep} = 1$. By obtaining the SINR value corresponding to BLER = $0.1$ from Figure~\ref{BLER}, the CQI mapping curve in Figure~\ref{mapping} can be derived after smoothing. The curves in Figure~\ref{BLER} represent the different MCSs' results of the NB-IoT link simulation obtained from the link-level simulator. As the MCS value increases, more information is transmitted using the same resources, deteriorating the BLER performance. A coin toss is used to determine whether the received transport block was decoded correctly or not. Precisely, an effective SINR $\gamma_{\rm eff}$ corresponds to a BLER value in Figure~\ref{BLER}. Subsequently, a uniformly distributed random number between $0$ and $1$ is generated. If the random number exceeds the BLER value, the decoding result is considered correct; otherwise, it is deemed incorrect.

\subsubsection{Feedback}
In the final step of the main simulation loop, the CQI obtained from Figure~\ref{mapping} provides feedback to the scheduler to indicate the channel quality. Similarly, the decoding result from the link performance model serves as HARQ information, which is then sent back to the scheduler. After an unavoidable delay, the scheduler receives the HARQ information and uses it to determine whether new data should be transmitted. Ensuring that the scheduler and feedback function are properly aligned is essential, enabling the scheduler to interpret the feedback accurately.

\subsection{Post-processing}
The initialization and pre-generation blocks' global parameters remain unchanged throughout the system-level simulation. The main simulation loop processes the data in the cycle of TTI, so the simulation results must be stored at the end of each TTI. When all the subframes are processed, the individual results are combined and averaged, followed by statistical analysis. The obtained values, ranging from throughput to SINR, are displayed and stored in a file. Curves about important parameters are plotted so that the performance of the system-level simulator is presented vividly.  

\section{Three Key Features Used in the Simulation}
\label{Section-Simulation}    
This section evaluates the maximum coupling loss (MCL), traffic model, and extended discontinuous reception (eDRX), which are imperative to the system-level simulation of an NB-IoT network. The MCL quantifies the wide-area coverage in an NB-IoT network. The traffic model theoretically analyzes the massive connection capability of NB-IoT terminals. Finally, eDRX verifies the low-power consumption of NB-IoT terminals.

\subsection{Maximum Coupling Loss}
In legacy cellular communication networks, MCL is widely used to reflect network coverage performance. By definition, MCL is the difference between transmit power and receiver sensitivity, given a receiver processing gain being $0$ dB. Receiver sensitivity is the sum of effective noise power and target signal-to-noise ratio (SNR). As shown in Table~\ref{MCL}, the MCL of downlink and uplink in the NB-IoT network are calculated to be $164$ dB. As for the link budget of the LTE technology, its MCL is designed as $144$ dB \cite{ratasuk2016nb}. Therefore, the NB-IoT network achieves a coverage enhancement of $20$ dB compared to legacy LTE.

\renewcommand\arraystretch{1.2}
\begin{table}[t]
\small
	\caption{Link budget for NB-IoT network.}
	\centering
	\begin{spacing}{1.0}
		\begin{tabular}{!{\vrule width1.2pt}p{4.6cm}!{\vrule width1.2pt}c!{\vrule width1.2pt}c!{\vrule width1.2pt}}
			\Xhline{1.2pt}
			{\bf Parameter} & {\bf downlink}   & {\bf uplink}              \\
			\Xhline{1.2pt}
			(1) Transmit power (dBm)               & 43  & 23        \\
			\hline
			(2) Thermal noise density (dBm/Hz)     & -174 & -174      \\
			\hline
			(3) Receiver noise figure (dB)         & 5   & 3          \\
			\hline
			(4) Interference margin (dB)           & 0   & 0           \\
			\hline
			(5) Occupied bandwidth (kHz)           &  180  & 15   \\
			\hline
			(6) Effective noise power  =  (2)  + (3) + (4) + 10$\log_{10}$((5)) (dBm) & -116.4  & -129.2     \\
			\hline
			(7) Target SNR (dB) & -4.6  & -11.8  \\
			\hline
			(8) Receiver sensitivity = (6) + (7) (dBm)& -121  & -141    \\
			\hline
			(9) Receiver process gain (dB) &  0 & 0  \\
			\hline
			(10) {MCL} = (1) - (8) + (9) (dB) & {164}  & {164}  \\ 
			\Xhline{1.2pt}
		\end{tabular}
	\end{spacing}
	\label{MCL} 
\end{table}

\subsection{Traffic Model}
\label{Traffic model} 
Various services are available in the NB-IoT network, and in this work, we focus on service clustering based on spatial-temporal characteristics. These services can be categorized into three groups: 
\begin{itemize}
	\item[i)] Services with fixed location and determined time, such as smart meters;
	\item[ii)] Services with fixed location but random time, like intelligent parking;
	\item[iii)] Services with random locations and times, such as shared bicycles.
\end{itemize}
Due to the diverse nature of NB-IoT services, it is impossible to represent them using a single traffic model. Hence, we analyze the traffic model based on the frequency and size of packets to assess the capability of handling massive connections.

Mobile autonomous reporting (MAR) measures the frequency of packets in the NB-IoT network. According to the periodic reports in MAR, the arrival period of packets is distributed over four categories with constant times of $1$ day, $2$ hours, $1$ hour, and $30$ minutes, with proportions of $40\%$, $40\%$, $15\%$ and $5\%$, respectively \cite{45820}. The average data rate of reports that a terminal should achieve, which is calculated based on the MAR periodic traffic model \cite{capacity}, is computed as follows:
\begin{align}
	R_{\rm av} &= {0.4}/{86400} + {0.4}/{7200} + {0.15}/{3600} + {0.05}/{1800} \nonumber \\ 
			 &= 129.6\times 10^{-6} \, {\rm packet/s/terminal}. \label{rate_data}
\end{align}
 
The size of packets in the NB-IoT network follows a Pareto distribution with a minimum application payload size (${\rm data}_{\rm{min}}$) of 24 bytes and a parameter $\kappa$ of $2.5$. Thus, the expectation of packet size is approximately $32$ bytes. By using \eqref{rate_data}, the average data rate can be calculated as:
\begin{align}
	R_{\rm av} &= 129.6\times 10^{-6} \times 32 \times 8 \nonumber \\
			 &= 33.2 \times 10^{-3} \, {\rm bit/s/terminal}.
\end{align}
Consequently, the total data rate of packets per second in a sector with $M$ terminals is computed as $R_{\rm tot} = M \times R_{\rm av}$. Based on the family density in London, the number of NB-IoT terminals a sector can support is $52549$ \cite{45820}. Therefore, to support such massive connections with $M = 52549$ NB-IoT terminals, $R_{\rm tot}$ can be given by
\begin{equation}
	R_{\rm tot} =  52549 R_{\rm av} = 1744.6 \,  {\rm bit/s/sector}.
\end{equation}

\begin{figure*}[!t]
	\begin{center}
		\includegraphics[width=5.25in, clip, keepaspectratio]{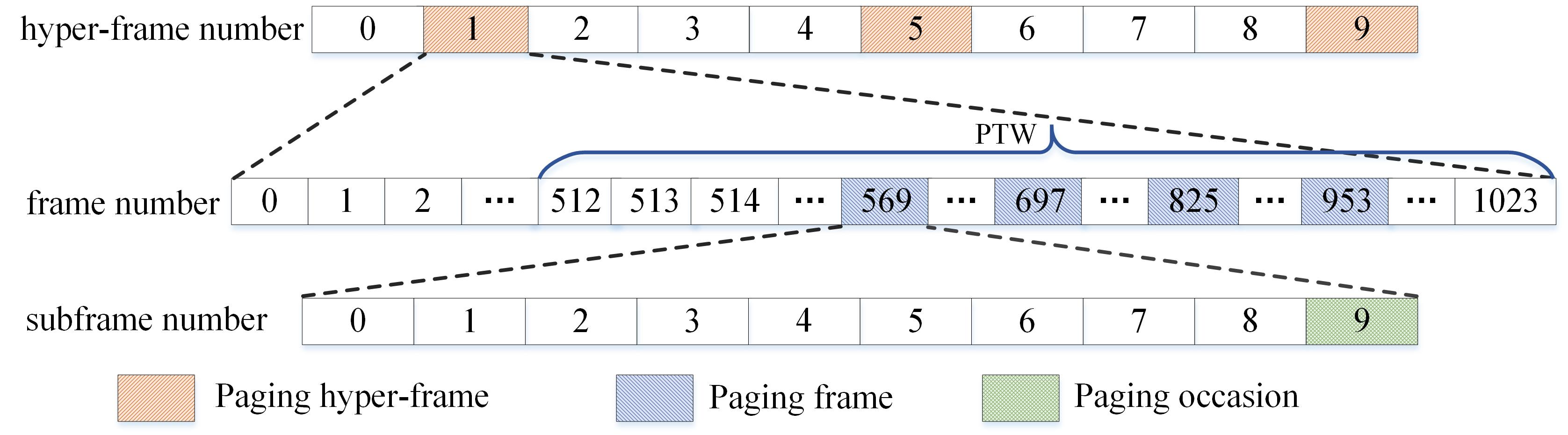}
		\caption{Function of eDRX in NB-IoT networks.}
		\label{PO}
	\end{center}
\end{figure*} 

\subsection{eDRX}
If a terminal continuously monitors the paging message from the network, it will result in high power consumption. To address this issue, 3GPP introduced the discontinuous reception (DRX) technique. Using DRX, a terminal periodically checks the paging message when paging may occur while it sleeps to save power for the rest of the time. However, the maximum period of the DRX is $2.56$ seconds, which does not meet the low-power consumption requirement of NB-IoT. Consequently, 3GPP proposed eDRX to further reduce the power consumption of NB-IoT terminals by extending the wake-up period.

The period of eDRX depends on a superframe, which has a length of $10.24$ seconds, comprising $1024$ frames. Each frame has $10$ subframes. There are $T_{\rm eDRX}=2^k$, where $k = 1, 2, 3, \cdots, 10$ superframes in an eDRX period, ranging from \{$20.48$, $40.96$, $81.92$, $163.84$, $327.68$, $655.36$, $1310.72$, $2621.44$, $5242.88$, $10485.76$\} seconds. The maximum period of the eDRX is up to $10485.76$ seconds, or equivalently $2.91$ hours, allowing a longer period for NB-IoT terminals. As shown in Figure~\ref{PO}, in eDRX, the superframe where the paging message is located is called the paging hyper-frame, the frame where the paging message is located is called the paging frame, and the subframe in which the paging message is located is called the paging occasion. The eDRX period includes the paging time window (PTW) to monitor paging messages and the idle period to sleep.

To implement the function of eDRX, we set up a paging timetable and timer between the NB-IoT eNodeB and NB-IoT terminal. According to the paging timetable, an NB-IoT terminal wakes up in PTW. The activated NB-IoT terminal associates with the NB-IoT network in case of a corresponding paging message. When a timeout occurs in the timer, the NB-IoT terminal returns to the idle period and rests.

\begin{figure}[!t]
	\begin{center}
		\includegraphics[width=3.0in]{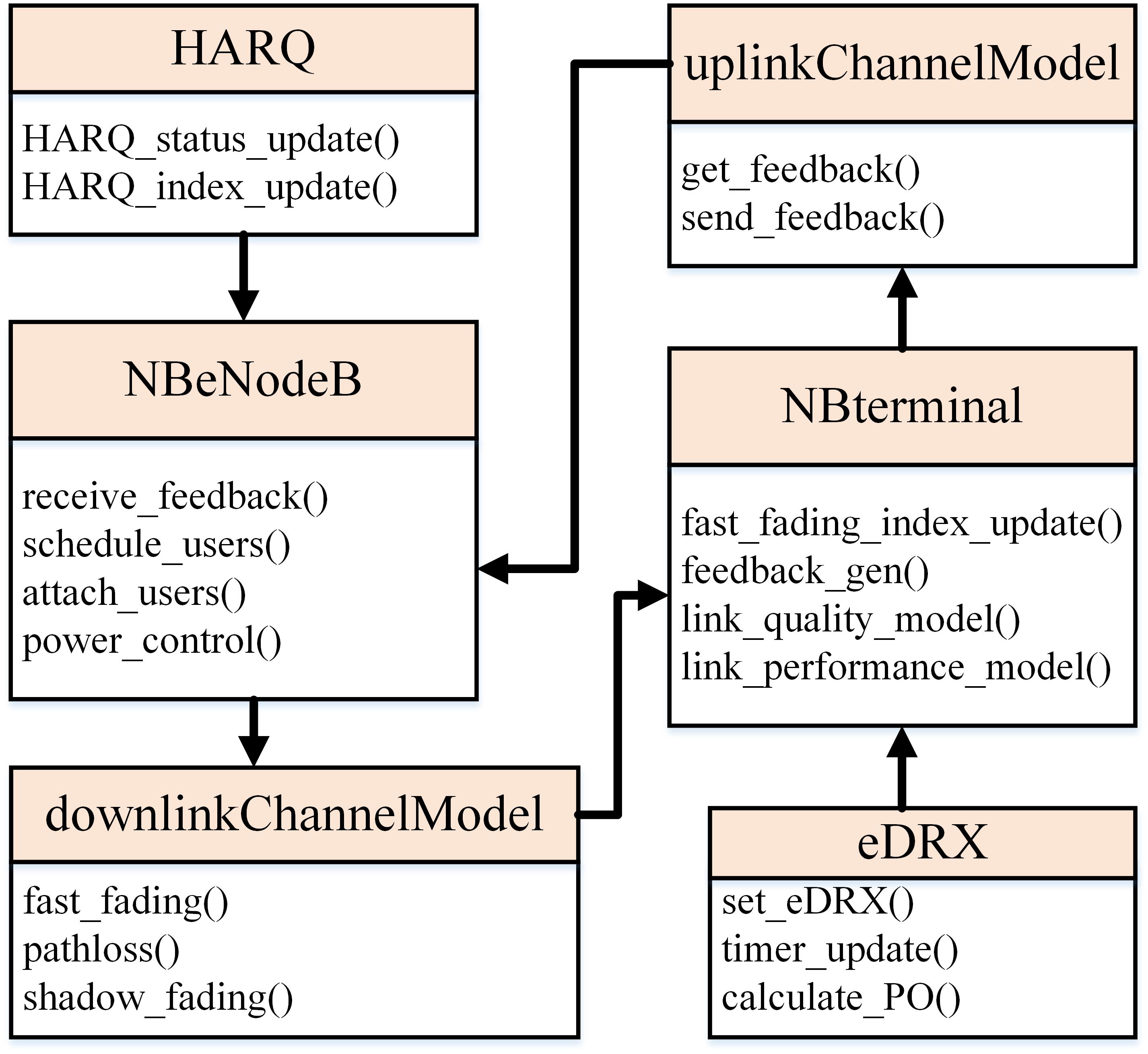}
		\caption{Class diagram of the system-level simulator.}
		\label{diagram}
	\end{center}
\end{figure}

\section{Simulation Results and Discussions}
\label{Section-Results}
In this section, we present and discuss Monte Carlo simulation results to verify the effectiveness of the developed system-level simulator. To speed up the simulation experiments, we model essential components  (e.g., NB-IoT eNodeB and NB-IoT terminal) as programming objects, depicted in Figure~\ref{diagram} with their implementation of public functions. The NB-IoT network is deployed in the stand-alone mode, and the related simulation parameters are summarized in Table~\ref{parameter}.  Both SINR and throughput are accounted for in simulation results.

\begin{table}[!t]
	\small
	\caption{Parameter setting for the system-level simulator.}
	\centering
	\begin{spacing}{1.0}
		\begin{tabular}{!{\vrule width1.2pt}c|c|c!{\vrule width1.2pt}}
			\Xhline{1.2pt}
			{No.} & {\bf Parameter}   & {\bf Value}              \\
			\Xhline{1.2pt}
			1   & Number of cells  & 19                        \\
			\hline
			2     & Inter-cell distance & $1732$ m             \\
			\hline
			3     & Center frequency    & $900$ MHz             \\
			\hline
			4  & Thermal noise density  & $-174$  dBm/Hz      \\
			\hline
			5  &  Max Tx power of downlink  & $43$ dBm            \\
			\hline
			6  &  Max Tx power  of uplink & $23$  dBm               \\
			\hline
			7  & std. deviation of shadowing  & $8$  dB         \\
			\hline
			8  & corr. distance of shadowing  & $110$  m       \\
			\hline
			9 & Downlink antenna gain  & $18$  dBi                      \\ 
			\hline
			10 & Uplink antenna  gain  & $-4$  dBi                         \\ 
			\Xhline{1.2pt}
		\end{tabular}
	\end{spacing}
	\label{parameter} 
\end{table}

\begin{figure}[t]
	\begin{center}
		\includegraphics[width=3.5in]{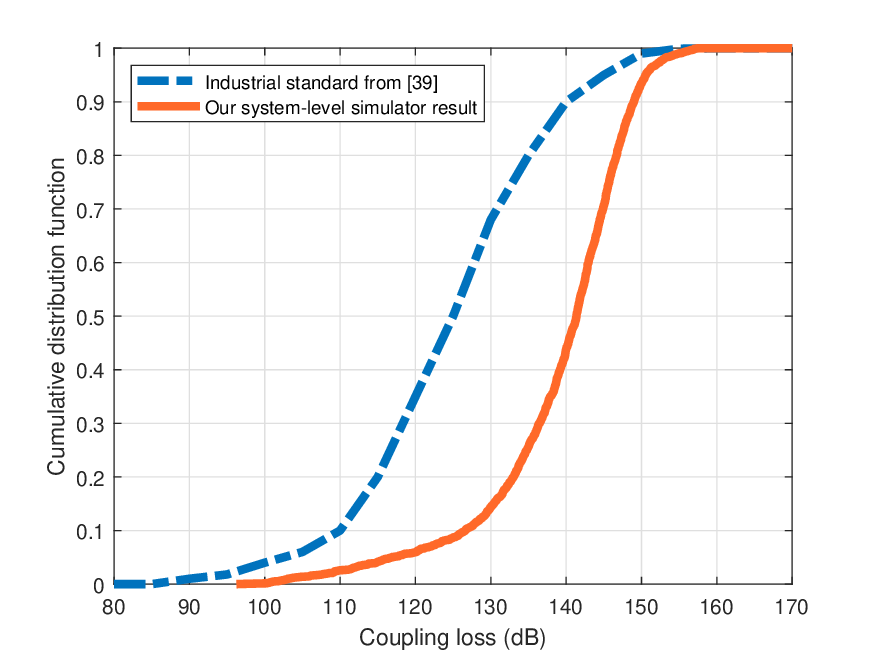}
		\caption{Comparison of system-level and industrial results w.r.t coupling loss.}
		\label{SINR}
	\end{center}
\end{figure}

Figure~\ref{SINR} plots the CDF curves of coupling loss calculated by SINR. By definition, the SINR between NB-IoT eNodeB and an NB-IoT terminal is the sum of transmit power, channel gain of large-scale fading, interference, and noise, all measured in \si{dB}. Our system-level simulation results are compared with the industrial results \cite{benchmark} reported by Ericsson in a 3GPP meeting. In our simulation, $43\%$ of the SINR values are less than $140$ dB, while this proportion is $90\%$ in the industrial results. This observation suggests that our simulation indicates a lower proportion of low SINR values, implying higher receiver sensitivity. Both curves increase and reach $100\%$ at $160$ \si{dB}, demonstrating that they meet the MCL requirement of $164$ dB in the NB-IoT network. However, our results exhibit a steeper slope.

\begin{figure}[t]
	\begin{center}
		\includegraphics[width=3.5in]{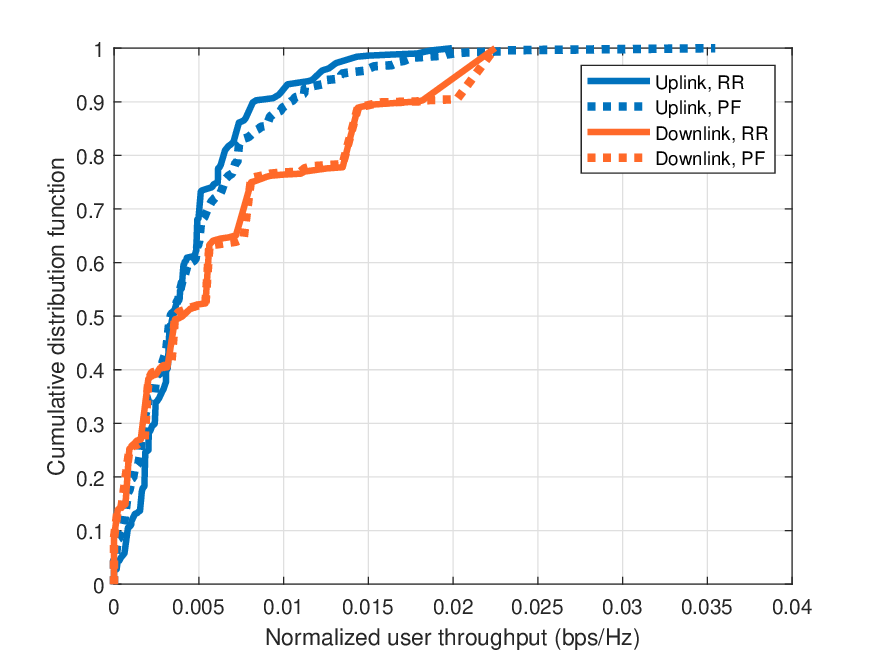}
		\caption{Comparison of different scheduling methods.}
		\label{totalSINR}
	\end{center}
\end{figure}

Figure~\ref{totalSINR} compares different scheduling strategies in terms of the normalized user throughput when using 4000 NB-IoT terminals in a sector employing eDRX. In this context, throughput refers to the transport block size divided by the transmission time and bandwidth, which is the number of bits transmitted within the unit bandwidth per unit time. The normalized user throughput is derived by dividing the throughput by the number of scheduled NB-IoT terminals. In the figure, the red curves represent downlink normalized throughputs, which are slightly larger than the blue curves, which represent uplink normalized user throughputs, regardless of whether RR or PF scheduling strategy is applied. This observation suggests that the NB-IoT downlink transmits more packets than the uplink. Furthermore, the figure also indicates that the maximum normalized user throughput in NB-IoT is approximately $0.02$ \si{bps/Hz}, notably smaller than the $3$ to $5$ \si{bps/Hz} typically achieved in legacy LTE systems. This is expected due to the much smaller packets in NB-IoT. Additionally, Figure~\ref{totalSINR} demonstrates that the RR and PF scheduling strategies deliver similar normalized user throughput in both downlink and uplink transmissions.

\begin{figure}[t]
	\begin{center}
		\includegraphics[width=3.5in]{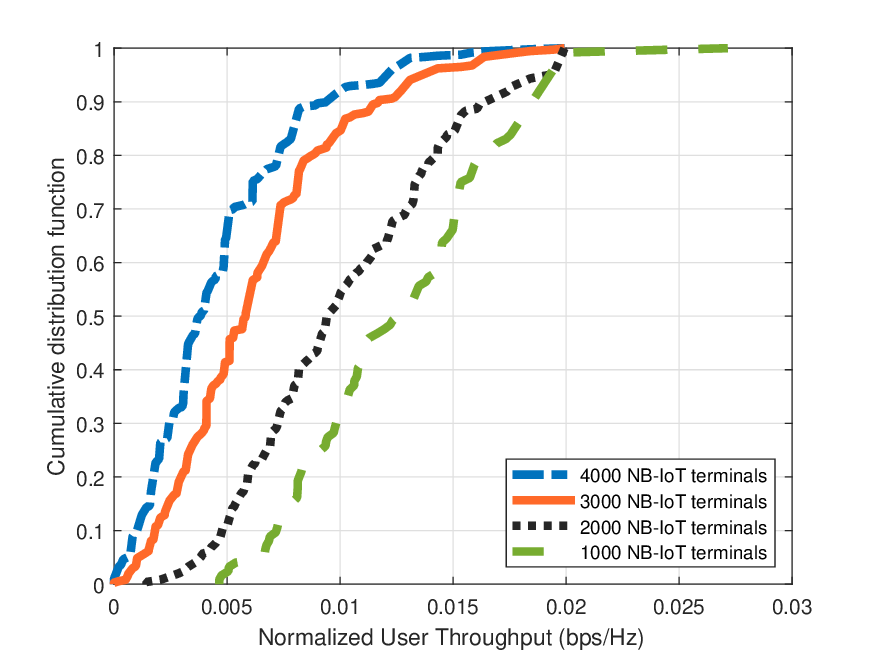}
		\caption{Comparison of the number of NB-IoT terminals.}
		\label{throughput}
	\end{center}
\end{figure}

Figure~\ref{throughput} illustrates the impact of the number of NB-IoT terminals on normalized user throughput when using RR scheduling and eDRX. The median normalized user throughput values for $4000$, $3000$, $2000$, and $1000$ NB-IoT terminals are $0.004$, $0.0056$, $0.0097$, and $0.013$ \si{bps/Hz}, respectively. It can be inferred from these results that the normalized user throughput increases as the number of NB-IoT terminals decreases. This is because, with fewer NB-IoT terminals, each terminal can be allocated more resources. Put simply, the normalized user throughput will be higher if the same amount of resources is shared among fewer NB-IoT terminals. Therefore, scheduling more NB-IoT terminals in the eDRX mode will take longer, which leads to a decrease in the normalized user throughput to accommodate more NB-IoT terminal connections.

\begin{figure}[t]
	\begin{center}
		\includegraphics[width=3.5in]{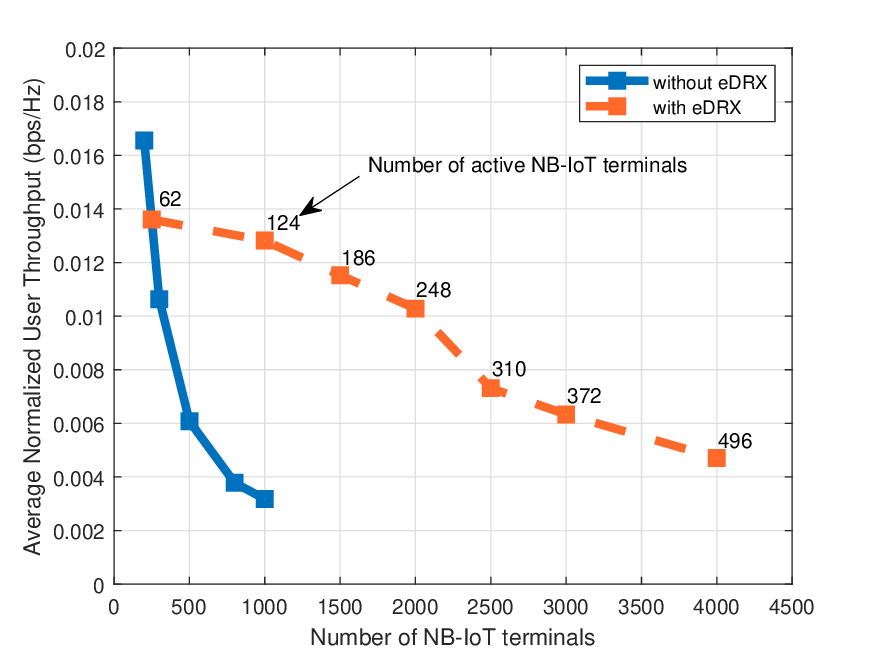}
		\caption{Comparison of eDRX with normalized user throughput.}
		\label{eDRX}
	\end{center}
\end{figure} 

Figure~\ref{eDRX} demonstrates the effect of eDRX on the average normalized user throughput. The average normalized user throughput is calculated by adding up all the elements in the normalized user throughput. As the number of NB-IoT terminals increases, the performance of average normalized user throughput without eDRX gradually deteriorates, which aligns with the conclusion in Figure~\ref{throughput}. When eDRX is used, not all NB-IoT terminals are active. For example, only $62$ out of $500$ NB-IoT terminals are activated, resulting in an average normalized user throughput of $0.0138$ \si{bps/Hz}. When the total number of NB-IoT terminals is $4000$ and eDRX is used, $496$ NB-IoT terminals are active simultaneously, and the corresponding average normalized user throughput is similar to that without eDRX. As expected, eDRX significantly improves the average normalized user throughput.

\section{Conclusions}
\label{Section-Conclusions}
This paper presented a practical system-level simulator for NB-IoT networks. The simulator is structured around four main building blocks: initialization, pre-generation, main simulation loop, and post-processing. It delved into three crucial aspects of NB-IoT networks: evaluation of MCL, traffic model, and eDRX. The simulation results effectively showcase the performance of SINR and throughput, providing a clear picture of how NB-IoT networks enable extensive connectivity. This practical system-level simulator is a reliable tool for evaluating performance and developing NB-IoT communication systems.

\bibliography{References}
\vfill
\end{document}